# Atmospheric aerosol clearing by femtosecond filaments

A. Goffin, J. Griff-McMahon, I. Larkin, and H.M. Milchberg*

*Institute for Research in Electronics and Applied Physics, University of Maryland, College Park, MD, 20742, USA*
*milch@umd.edu

Atmospheric aerosols, such as water droplets in fog, interfere with laser propagation through scattering and absorption. Femtosecond optical filaments have been shown to clear foggy regions, improving transmission of subsequent pulses. However, the detailed fog clearing mechanism had yet to be determined. Here we directly measure and simulate the dynamics of ~5 μm radius water droplets, typical of fog, under the influence of optical and acoustic interactions characteristic of femtosecond filaments. We find that for filaments generated by the collapse of collimated near-infrared femtosecond pulses, the main droplet clearing mechanism is optical shattering by laser light. For such filaments, the single cycle acoustic wave launched by filament energy deposition in air leaves droplets intact and drives negligible transverse displacement, and therefore negligible fog clearing. Only for tightly focused non-filamentary pulses, where local energy deposition greatly exceeds that of a filament, do acoustic waves significantly displace aerosols.

---

## I. INTRODUCTION

Filamentation of short powerful laser pulses in transparent optical media occurs when self-focusing collapse from the Kerr effect is arrested by ionization. For femtosecond laser pulses in air, the dynamic interplay between self-focusing and ionization defocusing gives rise to a self-guided beam consisting of a high intensity "core" region of diameter $d_\text{core} \sim 200 \mu\text{m}$ surrounded by a lower intensity "reservoir" which continuously exchanges energy with the core during propagation [1-3]. In air, the peak core intensity is limited by plasma refraction to $< \sim 10^{14}$ W/cm$^2$ [4]. Typically, the axial extent of intense core propagation is much greater than the Rayleigh range corresponding to $d_\text{core}$; this is of high interest to applications of long-range propagation [5,6]. The axial extent of the filament is approximately set by the overall diameter (and therefore Rayleigh range) of the filamenting beam [7]—for example, a 2 cm diameter beam of sufficient peak power can generate a filament over hundreds of meters. Filaments are resistant to small blockages due their core plus reservoir structure: if the high intensity core is obstructed by a water droplet or aerosol, it is reformed downstream by energy flow from the surrounding reservoir [8–10].

In air, filaments deposit energy primarily through plasma generation and rotational excitation of O$_2$ and N$_2$ [11,12]. Inverse Bremsstrahlung heating from electron-ion collisions is negligible for the sub-picosecond pulses typically used in air filamentation [13]. Energy deposition is impulsive on the timescale of the acoustic response of air to the filament, $\tau_a \sim d_\text{core}/2c_s \sim 300$ ns, using $c_s \sim 300$ m/s for the speed of sound in atmosphere. The impulsively heated air launches a locally cylindrical single-cycle acoustic wave, leaving behind a density depression, or "density hole", along the laser axis that lasts for several milliseconds [14,15]. The strength of the acoustic wave and depth of the density depression are each proportional to the energy deposited per unit length [7].

One important application of filamentation is fog clearing. Prior experiments [16,17] have shown that a filament propagating in a fog chamber improves the transmission of optical pulses injected immediately afterward along the filament axis. Fog droplet clearing has also been demonstrated using a timed sequence of pump pulses which heat the air by resonantly exciting molecular rotations [18]. Such droplet clearing may have applications in optical data transmission [16] or directed energy [19]. While the physical mechanism for improved transmission has been suggested to be droplet clearing by the laser-induced acoustic wave [17, 18], to date there have been no direct measurements of this or other possible mechanisms.

In this paper, we directly measure and simulate the dynamics of carefully positioned ~5 µm radius water droplets under the influence of optical and acoustic interactions characteristic of femtosecond filaments in air. Droplets of this size are well within the typical aerosol size distribution of fog [20,21]. We find that for filaments generated by the collapse of collimated near-infrared femtosecond pulses, the main clearing mechanism of droplets is optical shattering by the laser light. For such filaments, the single cycle acoustic wave launched by filament energy deposition leaves droplets intact and drives negligible radial displacement. We find that only for tightly focused non-filamentary pulses, where local energy deposition greatly exceeds that of a filament, can acoustic waves significantly displace aerosols.

## II. MECHANISMS FOR WATER DROPLET CLEARING

We anticipate two main droplet clearing mechanisms and set up two experiments to study them. We consider a droplet "cleared" when it is either transversely displaced, largely intact, from the beam path of a subsequent pulse, or shattered into sufficiently small fragments (in the Rayleigh scattering regime) that scattering losses of a subsequent pulse would be greatly reduced. For both experiments, we used a Ti:Sapphire ($\lambda_0$ = 812 nm, pulse full width at half maximum $\tau$ = 45 fs, pulse rate 10 Hz) pump beam for generating a filament and a spatially-filtered, frequency-doubled Nd:YAG ($\lambda$ = 532 nm, $\tau$ ~ 7 ns, 10 Hz) probe beam to image droplets and the hydrodynamics driven by the laser-heated air. The delay of the probe relative to the pump was electronically controlled by a digital delay box, with few nanosecond jitter insignificant compared to the displacement time of the droplet. Individual $5 \pm 1.5$ µm radius distilled water droplets were generated by a piezoelectric-driven 10 µm inner diameter nozzle mounted on a 3D translation stage and synchronized to the pump and probe pulses (see Fig. 1(a)). The nozzle was positioned above the laser axis and pointing down, so the droplets entered the laser beam by gravity. Rough positioning of the droplet with respect to the laser axis was controlled using the digital delay, with fine 2D positioning in the transverse plane controlled with the translation stage. The droplet was positioned at the longitudinal location where the filament energy deposition and acoustic wave amplitude were maximized.

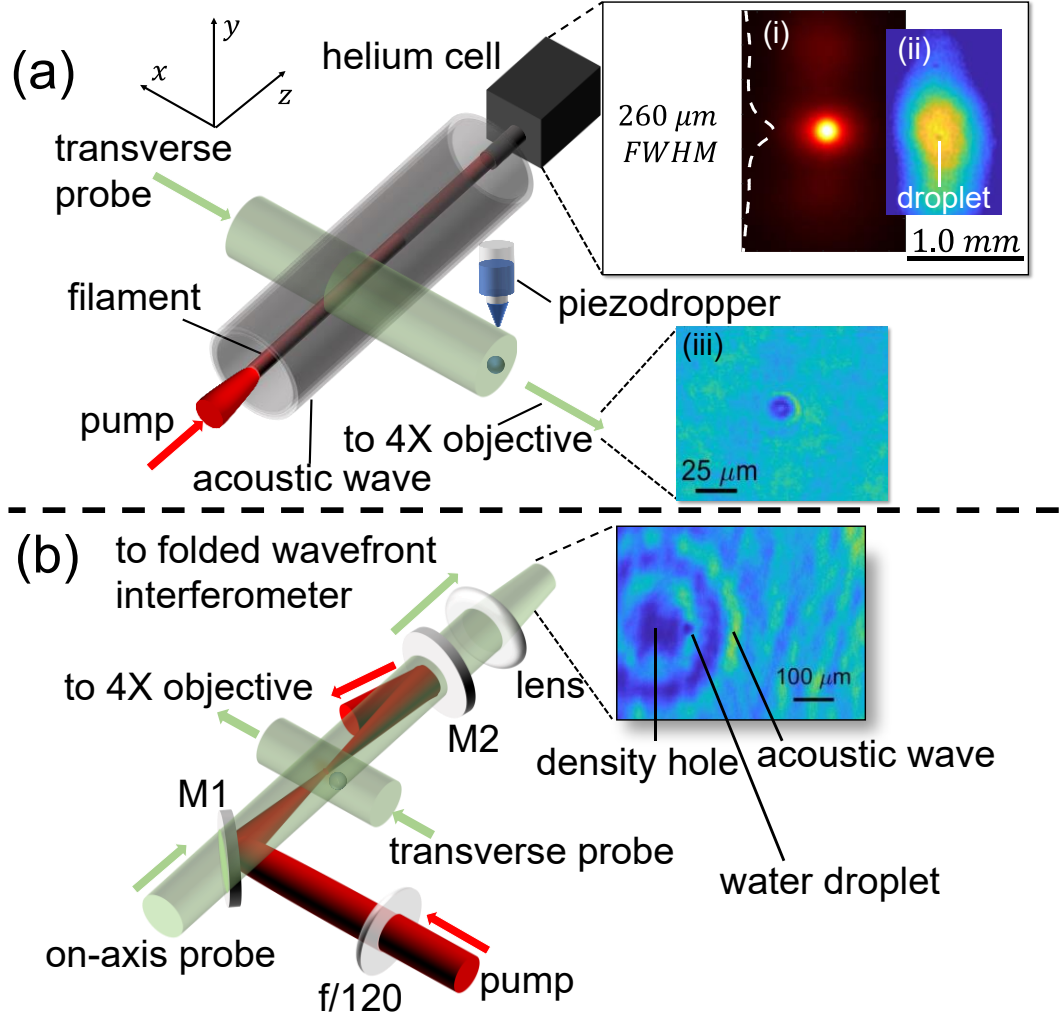

**Figure 1. (a)** Diagram of Experiment 1. A beam is down-collimated by a reflective telescope to a $w_0 = 1$ mm waist ($e^{-2}$ intensity radius) to generate a filament, radiating a single-cycle acoustic wave. The filament is terminated mid-flight by a helium cell, enabling linear imaging of the filament cross section [22,23]. A piezodropper places a 5 μm radius droplet a controlled distance from the center of the filament core; the droplet interaction is imaged from the side using a λ=532 nm, 7 ns transverse probe. (*i*) End-on image of the filament core, using the He cell, with the $y$ lineout plotted. (*ii*) End-on image of the droplet in a low-power beam. (*iii*) Side-image of an unperturbed 5μm radius droplet. **(b)** Diagram of Experiment 2. Two dielectric mirrors are used to co-propagate the pump and on-axis probe (M1), then filter out the pump to image the probe (M2). *Inset:* End-on image showing the density hole, radial acoustic wave, and water droplet.

In Experiment 1 (Fig. 1(a)), the pump beam is a filament generated using a near-collimated beam, where self-focusing is not assisted by a focusing optic (in contrast to [16,17]). The collimated beam configuration is most likely to be used for long range filament applications in the field. The goal in this experiment was to observe the relative contributions of radial droplet displacement and droplet shattering to clearing by a filament without lens assistance.

To generate single filaments in the lab, a 2.8 mJ pump pulse was down-collimated by a reflective telescope to a $w_0 = 1$ mm waist ($e^{-2}$ intensity radius). The filament was terminated mid-flight ~40cm after collapse and immediately after interaction with the droplet by an abrupt ~4 mm air-to-helium transition in the 1/2" diameter nozzle of a slowly outflowing helium cell [22,23], maintained at a slight positive pressure to the ambient air. Helium cell termination of the filament enabled end-on imaging of the filament intensity profile and calibration of the droplet position with respect to the center of the filament. The intensity profile was further attenuated by wedges in and after the helium cell to ensure linear propagation for imaging. Insets of Fig. 1(a) show images from the helium cell: an end-on image and lineout of a filament core (with $d_{\text{core}} \sim 260$ µm, in good agreement with propagation simulations for these conditions [24]), and an image of an unfilamented lower energy beam with a droplet 50 µm away from the beam axis. A side-imaging system with $4\times$ magnification (see Fig. 1(a) and (b)) used the transverse $\lambda = 532$ nm probe beam to image filament-droplet interaction dynamics.

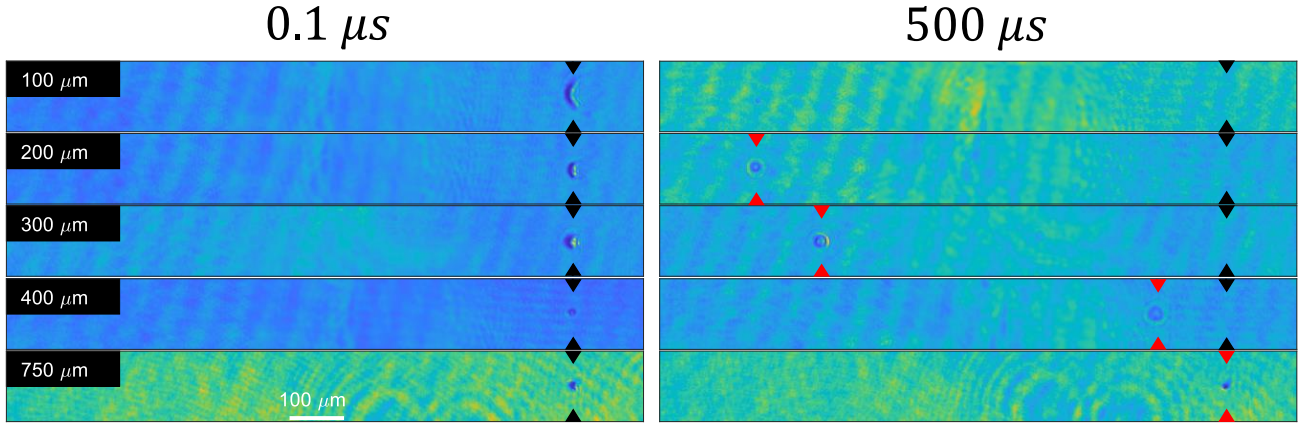

**Figure 2.** Shadowgrams of filament-induced dynamics for ~5 µm radius droplets placed at varying radial distance (labels at left) from the center of the filament core. The filament propagates from left to right (white arrows). The two columns show droplet images at 0.1 µs and 500 µs after filament arrival. The black arrows mark the initial droplet axial location, while the red arrows mark the droplet location (if a droplet is still present) 500 µs after the filament interaction. As the droplet is moved closer to the beam axis, the images at 0.1 µs show droplet distortion and far side cavitation, while the 500 µs images show axial droplet displacement toward the laser and the complete disintegration of droplets placed closer than ~200 µm from the core centre.

Figure 2 shows a sequence of side-shadowgrams at 0.1 µs and 500 µs after the filament of the droplet interaction as the droplet is translated closer to the filament core center from directly above. The 500 µs delay is chosen to match that for optimal transmission of a pulse following the filament in a prior fog clearing experiment [16]. It is seen from the 0.1 µs-delay panels that at ~750 µm from the center the droplet is unaffected, but as the droplet is moved closer, it increasingly deforms and cavitates owing to mass ejection from its far side (away from the laser). In the corresponding 500 µs delay panels, the droplet is seen to have reformed its shape and moved along the optical axis *toward* the laser, owing to a momentum boost from the mass ejection. At less than 200 µm (150 µm) from the core in the vertical (horizontal) direction, the droplet has completely disintegrated; in the 500 µs panel, there is no evidence of probe scattering by droplet fragments. The vertical-horizontal asymmetry originates from slight asymmetry of the filament cross section, as seen in Fig. 1(a)(i). At no droplet position is there any evidence of *radial* droplet displacement away from the filament, which is what one would expect if the cylindrical acoustic wave contributed significantly to droplet clearing [17]. Since droplet shattering occurs at the edge of the filament core but well inside the reservoir radius ($w_0 \sim 1$ mm), with cavitation and mass expulsion occurring on the far side of the droplet, we conclude that laser light from the filament reservoir is focused inside the droplet, with strong heating on the far side. This scenario has been discussed in [25], where the geometric focus of a droplet with $2\pi a/\lambda \gg 1$ (easily satisfied by our radius $a = 5$ μm droplets) leads to a radius-independent ~7.6× intensity enhancement at the far side, promoting cavitation and mass expulsion.

The goal of Experiment 2 was to observe droplet clearing effects using a laser intensity profile resembling the most intense part of the filament core (~100 μm, or approximately half the core diameter) where laser energy is deposited from plasma generation. The setup, shown in Fig. 1(b), uses an $f/120$ lens-focused pulse so that the beam waist diameter ($2w_0$ ~ 100 μm) matches this. With little to no laser flux radially outside $\sim w_0$, any droplet clearing from that region would not be directly optically-driven; a strong candidate would be clearing by the laser-induced acoustic wave. In the experiment, the pulse energy $\varepsilon_{pump}$ was adjusted in the range $100 - 500$ μJ in order to match and exceed the energy deposition per unit length, $\partial_z \varepsilon_{dep}$, of a real filament core, and thus launch an acoustic wave of at least equal amplitude. The energy deposition was measured interferometrically using a portion of the $\lambda = 532$ nm pulse transmitted through the $\lambda = 800$ nm mirror M1 and directed along the pump beam axis. This on-axis probe picks up a phase shift $\Delta\Phi(\mathbf{r}_\perp)$ from the long-lasting air density depression left by pump heating, as described in [7]. The energy deposition per unit length is given by [7] $\partial_z \varepsilon_{dep} = -c_v T_0 \rho_0 k^{-1} L^{-1} (n-1)^{-1} \int d^2 \mathbf{r}_\perp \Delta\Phi(\mathbf{r}_\perp)$, where the integral is over the cross section of the density hole, the simulated [24] axial FWHM of air heating is $L \sim 2$ cm (less than the confocal parameter of $2z_0 \sim 2.8$ cm owing to the strongly nonlinear heating), $c_v = 0.72$ kJ/(kg K) is the isochoric specific heat of air [26], $T_0 = 297$ K is the ambient air temperature, $\rho_0 = 1.23$ kg/m$^3$ is the air mass density [27], $k^{-1} = 84.7$ nm is the inverse wavenumber of the probe, and $(n-1) = 2.8 \times 10^{-4}$ is the air index of refraction increment [28].

Energy deposition $\partial_z \varepsilon_{dep}$ vs. $\varepsilon_{pump}$ is plotted in Fig. 3(a). For $\varepsilon_{pump} = 450$ μJ, $\partial_z \varepsilon_{dep}$ exceeds by ~40% our prior measurements of maximum deposition of 3.5 μJ/cm in a real filament [7]. Therefore, near the focus in Experiment 2, the launched acoustic wave for $\varepsilon_{pump} = 450$ μJ is stronger than an acoustic wave from a real filament; we use this experiment to determine an upper bound on acoustically-induced droplet displacement.

The droplet initial and final positions, $R_{init}$ and $R_{final}$, were imaged using the transverse $\lambda = 532$ nm, $\tau \sim 7$ ns probe pulse. To enable sensitive detection of potentially small laser-induced position changes, every other pump pulse was blocked for collection of a pump-off droplet image. These were then binned based on the droplet's blocked-pump position ($R_{init}$) with respect to the pump axis. The mean droplet displacement ($\overline{\Delta R}_{drop} = \bar{R}_{final} - \bar{R}_{init}$) after 500 μs delay is plotted in Fig. 3(b) as a function of $\varepsilon_{pump}$ for $R_{init}$ in the range $100 - 110$ μm, which encompasses the slight shot-to-shot variations in initial droplet location. The overbars represent a mean over ~100 shots at each point. For $\varepsilon_{pump} = 450$ μJ, the droplet is shattered by the beam focus for $R_{init} < 100$ μm. Figure 3(c) plots $\overline{\Delta R}_{drop}$ vs. $\bar{R}_{init}$ at 500 μs delay for $\varepsilon_{pump} = 450$ μJ. In both Fig 3(b) and (c), the error bars represent the displacement variance $\pm \Delta R_{var}$ over the ~100 shots at each point. The red bands in these panels are results from a hydrocode-simulation of acoustically-driven droplet displacement (to be discussed in Sec. III).

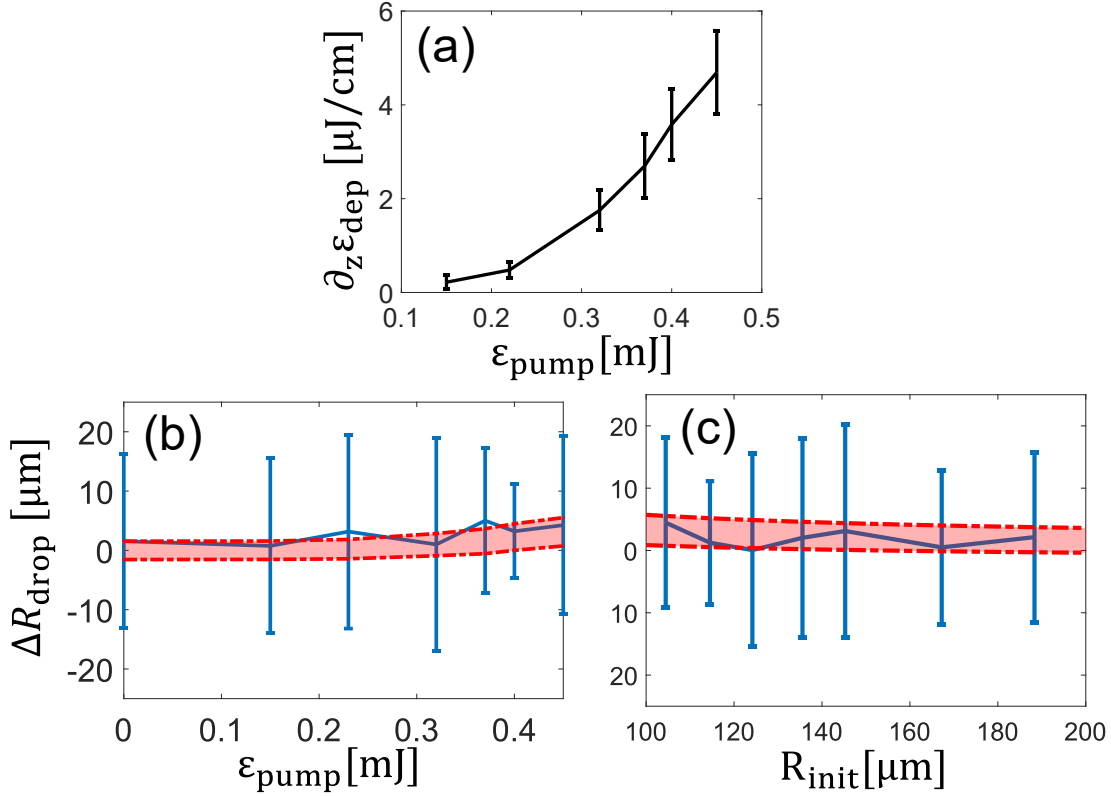

**Figure 3.** (a) Energy deposition per unit length $\partial_z \varepsilon_{dep}$ vs. pump energy $\varepsilon_{pump}$. (b) Droplet displacement $\Delta R_{drop}$ as a function of pump energy for initial position $R_{init} = 100\ \mu m - 110\ \mu m$. The blue trace is the mean displacement $\overline{\Delta R}_{drop}$ and the error bars represent variance over 100 shots at each point. The red region is $\Delta R_{drop}$ from simulations (see Sec. III). (c) $\Delta R_{drop}$ vs. $R_{init}$ for $\varepsilon_{pump} = 450\ \mu J$. Each value of $R_{init}$ plotted corresponds to the average at that point, with standard deviation $< 4.5\ \mu m$.

Both Fig. 3(b) and 3(c) show a very weak trend of increasing droplet displacement for increasing $\varepsilon_{pump}$ and decreasing $R_{init}$. In both cases, however, the maximum $\overline{\Delta R}_{drop}$ of ~5 μm is small compared to both $\Delta R_{var}$ and the variation of initial droplet positions. Even if we consider a maximum acoustic clearing displacement of $\Delta R_{var} \sim 20$ μm, this is still insufficient for acoustic droplet clearing by a real filament: for $R_{init} = 100$ μm, the displaced droplet would still be well within the filament's optical shattering radius (150-200 μm), as discussed earlier for Experiment 1. We therefore conclude that in a real filament, the intense core alone would not generate an acoustic wave of sufficient strength to clear a ~5 μm radius water droplet typical of fog.

## III. SIMULATIONS

Our experimental results for acoustic droplet clearing were compared with a 1D+time cylindrically symmetric hydrocode. Because energy deposition by a filament is impulsive [11,14], we assume that the initial excitation of the air is a heated region with radial temperature profile $\Delta T(r, t = 0) = \Delta T_{peak} \exp(-2(r/r_h)^2)$, where $r_h = 60$ μm (unless otherwise stated) and $\Delta T_{peak} = 2(\pi r_h^2 \rho_0 c_v)^{-1} \partial_z \varepsilon_{dep}$ is given by the measured energy deposition of Fig. 3, and plotted versus $\partial_z \varepsilon_{dep}$ in Fig. 4(a). The simulation then computes the gas evolution and the resulting drag and pressure gradient forces on the droplet.

The hydrocode solves (using [29]) the fluid conservation equations

$$\frac{\partial \xi_i}{\partial t} + \frac{1}{r}\frac{\partial}{\partial r}r(\xi_i v + \phi_i) = 0 \tag{1}$$

where the $\xi_i$ are the conserved densities and the $\phi_i$ are their forced fluxes. Here, $\xi_1 = \rho(r,t)$ is the air mass density, $\xi_2 = \rho v(r,t)$ is the momentum density, and $\xi_3 = \epsilon + \rho v^2/2$ is the energy density, where the air is taken as an ideal gas of internal energy density $\epsilon(r,t) = 3(\rho/m)k_B T(r,t)/2$ and pressure $P(r,t) = 2\epsilon/3$, with an average molecular mass $m = 0.8 m_{N2} + 0.2 m_{O2}$. The corresponding source fluxes are $\phi_1 = 0$, $\phi_2 = P + \tau$, and $\phi_3 = (P + \tau)v + q$, where $\tau = \frac{4}{3}\eta(T)\,\partial v/\partial r$ is the shear stress and $q = -\kappa(T)\,\partial T/\partial r$ is the thermal flux. Here, $\eta$ is the dynamic viscosity and $\kappa$ is the thermal conductivity of air [30].

The output of the fluid simulation is then used to compute the acoustically-driven motion of a spherical water droplet of radius $a$, mass $M$ and radial position $R$, where the force $F_{drop}$ on the droplet is taken as the sum of the local pressure gradient and drag force from the acoustic wave:

$$M\frac{d^2 R}{dt^2} = F_{drop} \tag{2}$$

$$F_{drop}(t) = \pi a^2 \left( \frac{\partial P}{\partial r} + 0.5 C_d \rho \left[\frac{dR}{dt} - v\right]^2 \mathrm{sgn}(v - \frac{dR}{dt}) \right)$$

Here, $p = p(R(t),t)$, $v = v(R(t),t)$, $\rho = \rho(R(t),t)$ are the fluid pressure, velocity, and density evaluated at the droplet position $R(t)$. The drag coefficient $C_d$ is a strong function of the Reynolds number ($Re = 2a\rho v\eta^{-1}$) of the flow around small droplets. We consider two drag regimes: $Re \leq 0.2$, for which $C_d = 24/Re$ follows Stokes' Law [30], and $Re > 0.2$, for which $C_d = 21.12/Re + 6.3/\sqrt{Re} + 0.25$ is a fit between the static $C_d = 0.47$ for $Re > 1000$ and Stokes' Law [30]. We note that $Re > 1000$ is unreachable with the droplet sizes and air/droplet velocities in this paper.

The simulation results are shown in the red bands overlaid on the experimental results in Figs. 3(b) and (c), showing excellent overlap, and verifying that filament-induced acoustic clearing of ~5 μm radius water droplets is negligible. The bands plot the simulated droplet displacement incorporating experimental shot-to-shot variations in energy deposition and initial droplet position $R_{init}$, and the uncertainty in droplet size. The energy deposition fluctuations were accounted for by initiating the hydrocode simulations by varying $\Delta T_{peak}$ for each $\varepsilon_{pump}$ in Fig. 3(a). The variation in droplet displacement owing to fluctuations in droplet radius $a$ was accounted for by simulating the smallest and largest $a$ within a standard deviation of the mean. Additionally, the mean displacement $\overline{\Delta R}_{drop}$ at $\varepsilon_{pump} = 0$ was added to the upper bound and subtracted from the lower bound of the simulated range since any acoustic displacement would add onto that detected (but not acoustic-induced) movement.

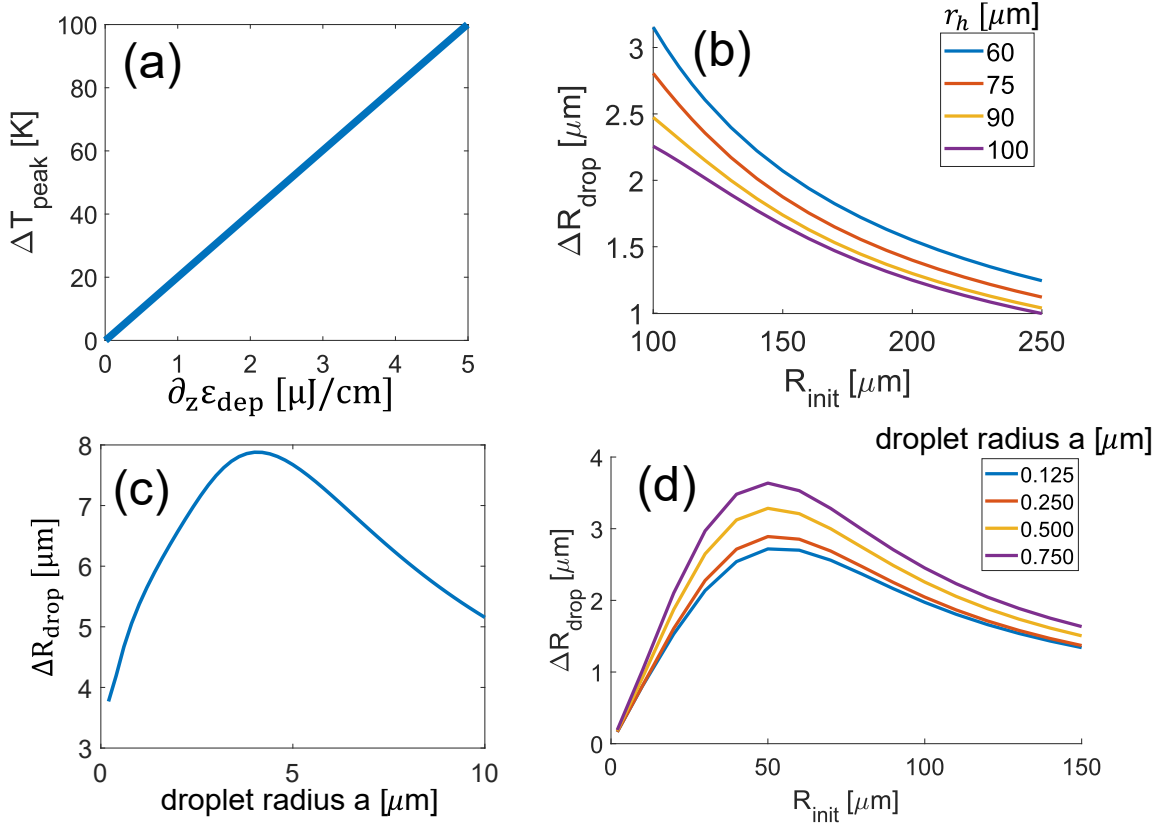

**Figure 4.** (a) Peak temperature rise vs. laser energy deposited per unit length $\partial_z \varepsilon_{dep}$. (b) Droplet displacement $\Delta R_{drop}$ at 500 µs delay for droplet radius $a = 5\ \mu m$, $\partial_z \varepsilon_{dep} = 4.8\ \mu J/cm$ for various $r_h$ (heated region radius) labelled in the legend. (c) $\Delta R_{drop}$ at 500 µs delay versus droplet radius $a$ for $\partial_z \varepsilon_{dep} = 4.8\ \mu J/cm$, initial droplet position $R_{init} = 50\ \mu m$, and $r_h = 60\ \mu m$. (d) $\Delta R_{drop}$ at 500 µs delay versus $R_{init}$ for $r_h = 60\ \mu m$ and droplet radii $a$ = 0.125 µm, 0.25 µm, 0.5 µm, 0.75 µm and $\partial_z \varepsilon_{dep} = 4.8\ \mu J/cm$. $\Delta R_{drop}$ decreases for $R_{init} < \sim 50\ \mu m$ because the acoustic wave reaches its maximum amplitude just inside the heated region.

To assess whether the radius of the filament-heated region can affect droplet movement, we performed additional simulations varying $r_h$, the $1/e^2$ radius of the initial temperature profile. A wider heated region may apply, for example, when laser energy is deposited through rotational excitations [11,13] outside of the highest intensity region of the filament core where optical field ionization primarily occurs. Simulation results plotted in Fig. 4(b) for an $a = 5$ µm droplet show that a fixed energy deposition $\partial_z \varepsilon_{dep}$ over a larger radius $r_h$ has reduced effect on droplet displacement, as one might expect from the reduced radial pressure gradient and lower peak fluid velocity. In all cases, the droplet displacement is negligible.

One possibility of interest is whether the acoustic wave can displace much smaller droplets, such as the fragments generated by optical shattering. Figure 4(c) plots droplet displacement as a function of droplet radius for $\partial_z \varepsilon_{dep} = 4.8\ \mu J/cm$ ($\Delta T_{peak} \sim 100$ K from Fig. 4(a)) and $R_{init} = 50$ µm (a near-in position where a droplet would normally be optically shattered), where it is seen that the greatest displacement is for droplets with radii in the ~5 µm range of our experiment. While smaller droplets accelerate more due to their smaller masses and larger drag coefficients, overriding the decrease in drag from their lower cross-sections, they also experience a stronger drag deceleration from the background air. This leads to the overall decrease of $\Delta R_{drop}$ as droplet radius decreases below ~5 µm. For droplet radii beyond ~5 µm, $\Delta R_{drop}$ decreases because of larger droplet inertia and reduced drag coefficient.

If the initial water droplet is shattered into fragments in a range of radii at different radial positions, Fig. 4(d) shows their possible displacements, neglecting the initial fragment velocity upon shattering (which is in all directions [31]). Here, the heated region radius is $r_h = 60$ µm. The smallest fragment simulated (radius $a = 0.125$ µm) is similar to the upper bound radius (0.135 µm) of laser-shattered water droplet fragments estimated in ref. [31]. For fragments of that size, $\Delta R_{drop}$ is negligible, and remains small for all fragments in Fig. 4(d), ranging to the larger droplets of Fig. 4(c). Therefore, we conclude that acoustic clearing is insignificant even for droplet fragments generated by optical shattering.

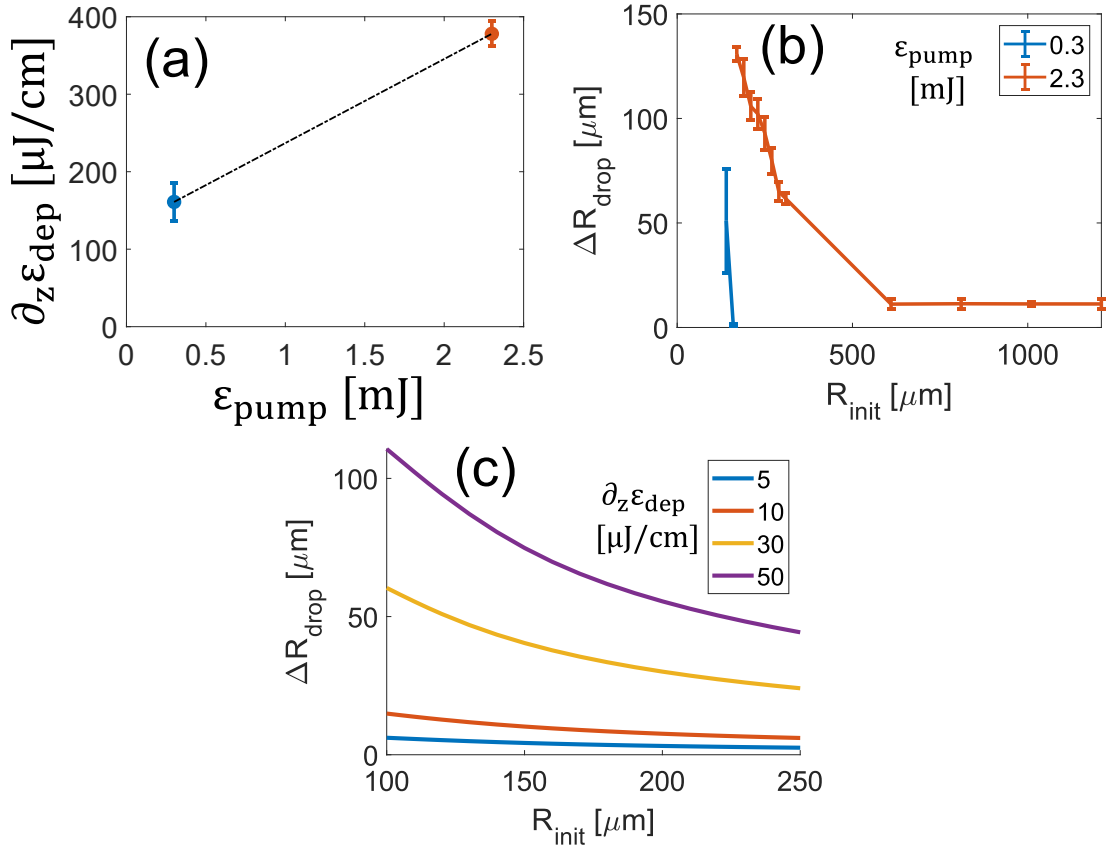

**Figure 5.** (a) Peak energy deposition per unit length for each pump pulse energy. For $\varepsilon_{pump} = 0.3$ mJ, $L \sim 1$ mm, and for $\varepsilon_{pump} = 2.3$ mJ, $L \sim 5$ mm [24]. (b) Droplet displacement for each pump energy after 105 µs delay. (c) Droplet displacement after 500 µs simulated by the hydrocode for $r_h = 60$ $\mu m$ and droplet diameter a ~ 10 µm, for the energy depositions labelled in the legend.

Although there is negligible acoustic clearing of water droplets from a collimated filament, there are non-filamentary conditions where acoustic clearing can be significant. We studied this by using a setup identical to Experiment 2 (Fig. 1(b)) except with $f/10$ focusing instead of $f/120$ and with larger water droplets of radius $a \sim 12.5$ µm. This setup produces a smaller heated region (for $\varepsilon_{pump} = 2.3$ mJ, simulation [24] gives $r_h \sim 45$ $\mu m$ and $L \sim 5$ mm) with significantly higher energy deposition and higher $\Delta T_{peak}$. In Fig. 5(a) the measured $\partial_z \varepsilon_{dep}$ is plotted for two values of $\varepsilon_{pump}$, with absorption of ~380 µJ/cm ($\Delta T_{peak} \sim 5000\ K$) at $\varepsilon_{pump} = 2.3$ mJ, roughly ~100 × higher than $\partial_z \varepsilon_{dep}$ in a filament. In Fig 5(b), the measured $\Delta R_{drop}$ at 105 µs delay is plotted vs. $R_{init}$. The displacement driven by a 2.3 mJ pulse is significant, with the droplet moving radially from ~170 µm to ~300 µm. For results more directly comparable to a filament, Fig. 5(c) plots results from the hydrocode using a $a = 5$ µm radius droplet and $r_h = 60\ \mu m$ as in Fig. 3, but higher $\partial_z \varepsilon_{dep}$. The results show that a heated region with the same spatial profile as a filament but with 10 × higher energy deposition can radially displace droplets significantly. For example, when $R_{init} = 150\ \mu m$ and $\partial_z \varepsilon_{dep} = 50$ µJ/cm, the droplet displacement is $\Delta R_{drop} = 75$ µm. However, in the deposition range for a filament ($\partial_z \varepsilon_{dep} < 5$ µJ/cm), droplet displacement is only $\Delta R_{drop}$ ~4 µm. This result shows that acoustic clearing is constrained mainly by the limited laser energy deposition in a near-IR femtosecond filament generated by collapse of a collimated beam.

We note that there may be self-guided propagation regimes in which acoustic clearing is more effective. For example, our recent simulations of self-guided propagation of 1 Terawatt, $\lambda \sim 10\ \mu m$, 3.5 ps pulses in air [32], show that peak values of $\partial_z \varepsilon_{dep}$ greater than ~100 µJ/cm are achievable from avalanche ionization and heating of ambient sub-micron aerosols by the self-guided pulse. This scenario may apply to recent experiments demonstrating self-guiding of long wavelength infrared pulses over tens of meters [33].

## IV. CONCLUSIONS

Self-guiding of short powerful pulses in the atmosphere has been proposed as a method for clearing atmospheric aerosols in fog for pulses that propagate after them, for applications including optical communications and directed energy. Our experiments and simulations show that for single filaments generated by collimated near-infrared femtosecond pulses, clearing of ~5 µm radius water droplets, typical of fog, is caused mainly by optical shattering for droplets closer than ~200 µm from the filament axis. Under these conditions, the filament reforms and continues to propagate and shatter droplets, clearing a path for a follow-on laser pulse or beam. "Clearing" in this case means reduced scattering losses of the follow-on beam owing to the much smaller radii of the shattered droplet fragments, a transition from Mie scattering to Rayleigh scattering. Droplets outside the shattering radius (>200 µm) undergo varying levels of distortion and cavitation as laser light is internally focused in the droplet, with heating and mass ejection on the far side driving the surviving portion of the droplet back toward the laser, but not cleared from the optical path. Detailed scattering losses from optically shattered droplets would need to be explored experimentally to reach a definitive conclusion on this method's clearing efficacy. We did not consider energy losses from droplet shattering, which by some estimates [34] would make fog clearing with a single near-IR filament over long distances impractical.

We find that acoustic forces are largely ineffective as a droplet clearing mechanism by near-infrared femtosecond filaments, mainly because the filament energy deposition of $< 5$ µJ/cm cannot drive an acoustic wave of sufficient amplitude to move a 5 µm droplet more than ~10 µm at best. Filament-induced acoustic waves are of little help clearing even the small droplet fragments generated by optical shattering—these experience offsetting drag forces by the ambient air.

Other fog clearing systems may be worth exploring. Long wavelength infrared (LWIR) filaments with few picosecond pulses [32,33] have large cross-sectional areas and high energies, and heat the air via avalanche ionization and collisional heating. This could lead to larger droplet shattering regions and greater energy deposition for acoustic clearing. LWIR pulses may also induce plasma-free droplet shattering [35], potentially decreasing laser energy losses. Finally, the resonant excitation of molecular rotations using pulse trains [12,18] can deposit energy per unit length exceeding that from plasma generation in a typical near-IR filament, increasing the acoustic wave amplitude and potentially enhancing clearing.

**ACKOWLEDGEMENTS**

The authors thank Andrew Tartaro, Aaron Schweinsberg, and Tony Valenzuela for technical assistance. This work is supported by the Air Force Office of Scientific Research and the JTO (FA9550-16-1-0121, FA9550-16-1-0284, and FA9550-21-1-0405), the Office of Naval Research (N00014-17-1-2705, N00014-20-1-2233), and the Army Research Lab (W911NF1620233).